\documentclass[a4paper,10pt]{article}
\usepackage{amsfonts, amsthm, amsmath}

\numberwithin{equation}{section}
\title{ {\bf
Second-order integrals for systems in $E_2$ involving spin}}

\author{\vspace{1cm}\\
         {\bf \.{I}smet Yurdu\c{s}en}
        \thanks{E-mail address:
       yurdusen@hacettepe.edu.tr}
\\
\\Department of Mathematics, Hacettepe University,
                     \\ 06800 Beytepe, Ankara, Turkey}

\date{\today}
\newpage
\begin{document}
\setlength{\baselineskip}{24pt} 
\maketitle
\setlength{\baselineskip}{7mm}
\begin{abstract}
In two-dimensional Euclidean plane, existence of second-order integrals of motion is investigated for integrable Hamiltonian systems involving spin (\emph{e.g.,} those systems describing interaction between two particles with spin 0 and spin 1/2) and it has been shown that no nontrivial second-order integrals of motion exist for such systems.
\end{abstract}


\newpage
\section{Introduction}
\label{intro}
In classical mechanics, existing of $n$ functionally independent integrals of motion defines integrability (in Liouville sense) of a Hamiltonian system with $n$ degrees of freedom. These integrals, including the Hamiltonian itself, must be well-defined functions on phase space and be in involution. This concept of integrability is extended to define superintegrability by requiring the existence of at least one and at most $n-1$ (in order to have dynamics in the system) additional integrals of motion. The total set of integrals of motion must be functionally independent, however, the additional  
ones are not necessarily in involution among themselves, nor with the already existing $n$ integrals of motion (except the Hamiltonian itself). All these concepts are also introduced in quantum mechanics through well-defined linear integrals of motion operators which are supposed to be algebraically independent \cite{Makarov, Evans.b, Tempesta}.

In quantum mechanics superintegrable systems are of physical interest because superintegrability entails exact solvability, meaning that the bound state energy levels can be calculated algebraically and the wave functions expressed in terms of polynomials in the appropriate variables, possibly multiplied by an overall factor. It has been conjectured \cite{Tempesta} that all maximally (having $2n-1$ integrals of motion) superintegrable systems are also exactly solvable and this has been supported by many examples \cite{Tempesta, Rodriguez}.

Systematic investigation of the superintegrable systems and their properties was initiated by the works of Smorodinsky, Winternitz and collaborators in 1965 \cite{Makarov, Evans.b}. Most of the earlier work was devoted on the quadratic superintegrability, (\emph{i.e.}, with integrals of motion that are second-order polynomials in the momenta) and directly related with the multiseparability in 2- and 3-dimensional Euclidean spaces \cite{Makarov, Evans.b}. Recently an extended review article has been published describing the current status of the subject \cite{rewMPW}.

Superintegrability properties are also investigated for Hamiltonian systems involving particles with spin \cite{WinYur1, FTY, wy3, DWY2012, Nikitin.d, Nikitin.f}. Systematic search for superintegrable systems with spin was initiated in \cite{WinYur1}, where the authors considered two nonrelativistic quantum particles, one with spin $\frac{1}{2}$, the other with spin $0$. Physically the most interesting Hamiltonian for such systems is 
\begin{eqnarray}
H=-\frac{\hbar^2}{2m} \Delta + V_0(\vec{r}) + \frac{1}{2}\Big\{V_1(\vec{r}), (\vec{\sigma}, \vec{L})\Big\}\,,
\label{sec1inteqH}
\end{eqnarray}
where $\{,\}$ denotes an anticommutator and $\sigma_1$, $\sigma_2$, 
$\sigma_3$ are the usual Pauli matrices. $V_0(\vec{r})$ and $V_1(\vec{r})$ are scalar and interaction potentials, respectively. This Hamiltonian given in $E_3$ would describe, for instance a low energy (nonrelativistic) pion--nucleon interaction. 

In \cite{WinYur1} first-order integrability and superintegrability was studied in $E_2$. Articles \cite{wy3, DWY2012} were devoted to systematic search of first- and second-order superintegrability in $E_3$.  

In this paper we will consider the Hamiltonian (\ref{sec1inteqH}) in $E_2$ and investigate the existence of second-order integrals of motion in order to classify further the superintegrable systems with spin in $E_2$. For simplicity we shall set the reduced mass $m$ of the two particle system equal to $m=1$ and use units in which the Planck constant is $\hbar=1$. Keeping $\hbar$ in the Hamiltonian and integrals of motion does not change any of the conclusions. In particular $V_0$ and $V_1$ do not depend on $\hbar$. In (\ref{sec1inteqH}) $H$ is a matrix operator acting on a two-component spinor and we will decompose it in terms of the $2 \times 2$ identity matrix $I$ and $\sigma_3$ (the matrix $I$ will be dropped whenever this does not cause confusion). $L$ is the angular momentum operator. 

In the next section we give the first-order integrable and superintegrable Hamiltonian systems, obtained from the analysis of the commutativity condition $[H, X_1] = 0$, where $X_1$ is the general first-order integral of motion in $E_2$. For details see \cite{WinYur1}. In Section 3, we search for the existence of second-order integrals of motion in $E_2$ for the two integrable cases obtained in Section 2. Finally, in the last section we give some conclusions.

\section{First-order integrability and superintegrability in $E_2$}\label{sec2}
In this section let us briefly review the results obtained in \cite{WinYur1}. 
Considering that the motion is constrained to a Euclidean plane (\emph{i.e.}, assuming $\Psi(\vec{r})=\Psi(x,y)$ and setting $p_3=0$, $z=0$), we have the following 
Hamiltonian
\begin{equation}
H = \frac{1}{2}({p_1}^2+{p_2}^2)+V_0(x,y)+V_1(x,y){\sigma}_3 L_3 +
\frac{1}{2}{\sigma}_3(L_3 V_1(x,y)) 
\label{sec2eq1}
\end{equation}
with 
\begin{eqnarray}
p_1=-i\partial_x, \qquad p_2=-i\partial_y, \qquad L_3=i(y\partial_x-x\partial_y), \qquad \sigma_3=\left(\begin{array}{cc} 1& 0\\ 0& -1 \end{array}\right)\,. \nonumber
\end{eqnarray}
The general first-order integral of motion to consider would be,
\begin{equation}
X_1 = \frac{1}{2}\{\mathcal{F}_1\,, p_1\} + \frac{1}{2}\{\mathcal{F}_2\,, p_2\} + \mathcal{F}_3\,, 
\label{integralofmotion}
\end{equation}
where 
\begin{equation}
\mathcal{F}_{\mu} = \mathcal{F}_{\mu 0} + \mathcal{F}_{\mu 1}\, \sigma_3\,, \qquad  \mu = 1,2,3\,.
\end{equation}
All six functions $\mathcal{F}_{\mu \nu}$, ($\mu = 1,2,3$ and $\nu = 0,1$) are real functions of $x$ and $y$. Our aim is to find at least one such integral of motion from the analysis of the commutativity condition $[H,X_1] = 0$. This condition provides $12$ determining equations for the $6$ functions $\mathcal{F}_{\mu \nu}$, as well as the unknown potentials $V_0(x,y)$ and $V_1(x,y)$. Six of the determining equations, which are obtained from equating the coefficients of the second-order terms to zero, give 
\begin{eqnarray}
&\mathcal{F}_{1\nu} = \omega_{\nu} y + a_{\nu}, \qquad \mathcal{F}_{2\nu} = -\omega_{\nu}x + b_{\nu}
\,, \qquad \nu = 0,1\,,
\label{sec21deteqs1}
\end{eqnarray}
where $\omega_{\nu}$, $a_{\nu}$ and $b_{\nu}$ are real constants 
and the rest of the determining equations are 
\begin{eqnarray}
&\mathcal{F}_{3\nu, x} = \delta_{\nu, 1-\xi}[-b_{\xi}V_1-(\omega_{\xi}y+a_{\xi})y 
V_{1,x}+(\omega_{\xi}x-b_{\xi})y V_{1,y}]\,, \nonumber \\
&\mathcal{F}_{3\nu, y} = \delta_{\nu, 1-\xi}[a_{\xi}V_1+(\omega_{\xi}y+a_{\xi})x 
V_{1,x}-(\omega_{\xi}x-b_{\xi})x V_{1,y}]\,, \nonumber \\
&(\omega_{\nu}y+a_{\nu})V_{0,x}+(-\omega_{\nu}x+b_{\nu})V_{0,y}=
\delta_{\nu, 1-\xi}(x\mathcal{F}_{3\xi, y} - y\mathcal{F}_{3\xi, x})V_1\,, \qquad (\nu, \xi) = (0,1)\,.
\label{sec21deteqs}
\end{eqnarray}

The analysis of the determining equations (\ref{sec21deteqs}) are summarized as: 

\noindent
{\bf 1) Superintegrable system}

There exists only one first-order superintegrable system with $V_1\neq0$:
\begin{equation}
H=-\frac{1}{2}\Delta + \frac{1}{2}{\gamma}^2(x^2+y^2) + \gamma \sigma_3 L_3,
\qquad \gamma=\mathrm{const}
\,. \label{firstordersupintH}
\end{equation}
It allows an $8$-dimensional Lie algebra $\mathcal{L}$ of first-order 
integrals of motion with a basis given by 
\begin{eqnarray}
L_{\pm} = i(y\partial_x-x\partial_y)I_{\pm}\,,\quad
X_{\pm} = (i\partial_x \mp \gamma y )I_{\pm}\,,\quad
Y_{\pm} = (i\partial_y \pm \gamma x )I_{\pm}\,,\quad
I_{\pm} = I\pm \sigma_3\,. 
\label{sec2algebra1}
\end{eqnarray}
The algebra $\mathcal{L}$ is isomorphic to the direct sum of two central 
extensions of the Euclidean Lie algebra $e(2)$
\begin{eqnarray}
\mathcal{L} \sim \tilde {e}_+(2)\oplus\tilde {e}_-(2)\,, \qquad
\tilde {e}_{\pm}(2)=\{L_{\pm}, X_{\pm}, Y_{\pm}, I_{\pm}\}
\,. \nonumber 
\end{eqnarray}

\noindent
{\bf 2) Integrable systems}

The only integrable systems with one integral of motion in addition to $H$ that we found are given as

\noindent
{\bf a) Integrable system with rotationally invariant potentials:}
 \begin{eqnarray}
V_0=V_0(r)\,, \qquad V_1=V_1(r)\,, \qquad r=\sqrt{x^2+y^2} 
\,, \nonumber \\
X =(\omega_0 + \omega_1 \sigma_3)L_3\,, \qquad \omega_{\nu}=\mathrm{const}
\,,\qquad \nu=0,1
\,. \label{rotinvsys}
\end{eqnarray}

\noindent
{\bf b) Integrable system with a $x$-dependent interaction potential:}
\begin{eqnarray}
V_1=V_1(x)\,, &\,& \qquad V_0=\frac{y^2}{2} {V_1}^2(x) + F(x)\,, \nonumber \\
X&=&-i\partial_y - \sigma_3 \int V_1(x)dx
\,. \label{xdeppotsys}
\end{eqnarray}

The above results should be understood up to gauge transformations of the form
\begin{equation}
\tilde H=U^{-1}HU, \qquad U=\left(\begin{array}{cc} e^{i\alpha}& 0\\
0& e^{-i\alpha} \end{array}\right), \qquad \alpha=\alpha(\kappa), 
\qquad \kappa = \frac{y}{x}
\,, \nonumber 
\end{equation}
which leaves the Hamiltonian (\ref{sec2eq1}) form invariant. However, the potentials transform accordingly
\begin{eqnarray}
\tilde {V_1}=V_1 + \frac{\dot {\alpha}}{x^2}, \qquad 
\tilde {V_0}=V_0+(1+\frac{y^2}{x^2})(\frac{1}{2}\frac{{\dot {\alpha}}^2}{x^2}
+ \dot {\alpha}V_1)
\,. \label{sectranspotsy}
\end{eqnarray}

\section{Second-order superintegrability in $E_2$}\label{sec3}
The system obtained in Section 2 with a constant interaction potential term is maximally superintegrable and hence all the higher-order integrals of motion can be expressed in terms of the first-order ones, given in (\ref{sec2algebra1}). However, for the integrable systems (\ref{rotinvsys}) and (\ref{xdeppotsys}) it is worth to search for the existence of second-order integrals of motion in order to classify further the superintegrable systems with spin in $E_2$. In this section, we investigate the existence of such second-order integrals of motion. 

\subsection{The potentials $V_0=V_0(r)$ and $V_1=V_1(r)$}
For these rotationally invariant potentials the Hamiltonian (\ref{sec2eq1}) becomes, 
\begin{equation}
H=\frac{1}{2}({p_1}^2+{p_2}^2)+V_0 + V_1\,\sigma_3\, L_3\,,
\label{sec3H}
\end{equation} 
where $V_0$ and $V_1$ are functions of $x^2 + y^2$ . The general second-order integral of motion to consider would be,
\begin{equation}
\tilde{X}_2 = \frac{1}{2}\{\mathcal{A}_1\,, p_1^2\} + \{\mathcal{A}_2\,, p_1p_2\} + \frac{1}{2}\{\mathcal{A}_3\,, p_2^2\} + X_1\,, 
\label{generalintegralv3}
\end{equation}
where 
\begin{equation}
\mathcal{A}_{\mu} = \mathcal{A}_{\mu 0} + \mathcal{A}_{\mu 1}\, \sigma_3\,, \qquad  \mu = 1,2,3\,,
\end{equation}
and $X_1$ is the first-order integral of motion given in (\ref{integralofmotion}). All six functions $\mathcal{A}_{\mu \nu}$, ($\mu = 1,2,3$ and $\nu = 0,1$) are real functions of $x$ and $y$. From the commutativity condition $[H, \tilde{X}_2] = 0$, we search for the existence of second-order integrals of motion. The highest-order determining equations (\emph{i.e.,} the determining equations, obtained by equating the coefficients of the third-order terms to zero in the commutativity equation $[H, \tilde{X}_2] = 0$) read:
\begin{eqnarray}
\mathcal{A}_{1 \nu\,,x} = 0\,, \quad \mathcal{A}_{3 \nu\,,y} = 0\,, \quad \mathcal{A}_{1 \nu\,,y} + 2  \mathcal{A}_{2 \nu\,,x}   = 0\,, \quad 2 \mathcal{A}_{2 \nu\,,y} +  \mathcal{A}_{3 \nu\,,x}   = 0\,, \quad (\nu=0,1)\,. 
\label{thirdorderdeterminingeqv3}
\end{eqnarray}
First four of the above equations imply $\mathcal{A}_{1 \nu} $ is a function of $y$ only and $\mathcal{A}_{3 \nu} $ is a function of $x$ only. Then the last four determining equations in (\ref{thirdorderdeterminingeqv3}) give 
\begin{eqnarray}
\mathcal{A}_{2 \nu\,,xx} = 0\,, \quad \mathcal{A}_{2 \nu\,,yy} = 0\,, \quad (\nu=0,1)\,, 
\label{solutionsthirdorderdeterminingeqv3}
\end{eqnarray}
which can immediately be integrated. Hence the general second-order integral of motion can be expressed as 
\begin{eqnarray}
X_2 = \mathcal{G}_1 (L_3\,p_1+p_1\,L_3) + \mathcal{G}_2 (L_3\,p_2+p_2\,L_3) + \mathcal{G}_3 ({p_1}^2-{p_2}^2) + 2\, \mathcal{G}_4\,p_1\,p_2 + X_1\,,
\label{secondorderintmotion}
\end{eqnarray}
where $\mathcal{G}_{\tau}$ ($\tau = 1, \dots, 4$) are now constants and all the determining equations, obtained by equating the coefficients of the third-order terms to zero in the commutativity equation are trivially satisfied. Since the potentials are rotationally invariant, the term proportional to $L_3^2$ is absent in (\ref{secondorderintmotion}) (\emph{i.e.,} it commutes with the Hamiltonian given in (\ref{sec3H})). Notice that the constants $\mathcal{G}_{\tau}$ ($\tau = 1, \dots, 4$) are considered as
\begin{equation}
\mathcal{G}_{\tau} = \mathcal{G}_{\tau 0} + \mathcal{G}_{\tau 1}\, \sigma_3\,, \quad  \tau = 1,\ldots,4\,, \qquad \mathcal{F}_{\mu} = \mathcal{F}_{\mu 0} + \mathcal{F}_{\mu 1}\, \sigma_3\,, \quad  \mu = 1,2,3\,.
\end{equation}

The determining equations, obtained by equating the coefficients
of the second-order terms to zero in the commutativity equation $[H, X_2] = 0$, read  
\begin{eqnarray}
&2\,\sigma_3\, \big((\mathcal{G}_4 - \mathcal{G}_2 y) V_1(x^2+y^2) + 2\,y\,\left [\mathcal{G}_3 x + y\,(\mathcal{G}_4 - \mathcal{G}_1 x - \mathcal{G}_2 y)\right]{V_1}^\prime (x^2+y^2) \big) + \mathcal{F}_{1,x} = 0\,, \label{sec2det1} \\
&2\,\sigma_3\, \big((\mathcal{G}_4 + \mathcal{G}_1 x) V_1(x^2+y^2) + 2\,x\,\left [-\mathcal{G}_3 y + x\,(\mathcal{G}_4 + \mathcal{G}_1 x + \mathcal{G}_2 y)\right]{V_1}^\prime (x^2+y^2) \big) - \mathcal{F}_{2,y} = 0\,, \label{sec2det2} \\
&2\sigma_3 \big((\mathcal{G}_2 x + \mathcal{G}_1 y - 2 \mathcal{G}_3) V_1(x^2+y^2) - 2\left[\mathcal{G}_3 (x^2+y^2) -2 x y(\mathcal{G}_1 x + \mathcal{G}_2 y)\right] {V_1}^\prime (x^2+y^2) \big) + \mathcal{F}_{1,y} + \mathcal{F}_{2,x} = 0,
\label{sec2det3}
\end{eqnarray}
where $^\prime$ denotes the derivative with respect to the argument. From equations (\ref{sec2det1})--(\ref{sec2det3}) we obtain compatibility conditions for $\mathcal{F}_{1\nu}$ and $\mathcal{F}_{2\nu}$ ($\nu = 0,1$), which are in polar form expressed as   
\begin{eqnarray}
\Gamma_{\nu} \left(r^3 V_{1,rrr} + 7r^2 V_{1,rr} + 9r V_{1,r} \right) + \Lambda_{\nu}\left(r^2 V_{1,rrr} + 3r V_{1,rr} - 3 V_{1,r} \right) = 0\,, \qquad \nu = 0,1\,, 
\label{sec3polarcom1}
\end{eqnarray}
where
\begin{equation}
\Gamma_{\nu} = \mathcal{G}_{1\nu} \cos \theta + \mathcal{G}_{2\nu} \sin \theta\,, \qquad \Lambda_{\nu} = \mathcal{G}_{4\nu} \cos 2\theta - \mathcal{G}_{3\nu} \sin 2\theta\,, \qquad \nu = 0,1\,.
\end{equation}

In general, (\ref{sec3polarcom1}) represents an overdetermined system of two different equations for the potentials $V_1$, namely
\begin{eqnarray}
&r^3 V_{1,rrr} + 7r^2 V_{1,rr} + 9r V_{1,r} = 0\,, \label{simultaneous1} \\
&r^2 V_{1,rrr} + 3r V_{1,rr} - 3 V_{1,r} = 0\,, 
\label{simultaneous2}
\end{eqnarray}
simultaneous solutions of which give 
\begin{equation}
V_1(r)=-\frac{\gamma_1}{2r^2} + \gamma_2\,, 
\label{casezero}
\end{equation}
where $\gamma_1$ and $\gamma_2$ are constants. Comparing (\ref{casezero}) with (\ref{sectranspotsy}) we see that we can cancel the constant $\gamma_1$ by a gauge transformation. Hence, we have a constant spin-orbit interaction potential that is found in Section \ref{sec2}. This system is maximally first-order superintegrable and thus all the higher-order integrals of motion can be expressed in terms of the first-order ones given in (\ref{sec2algebra1}).

When $\Gamma_0 = \lambda \Gamma_1$ and $\Lambda_0 = \lambda \Lambda_1$ 
with $\lambda = $ constant, an exception occurs and the two equations (\ref{sec3polarcom1}) coincide. Hence, bearing in mind that $V_1$ does not depend on $\theta$, now the equation (\ref{sec3polarcom1}) implies either (\ref{simultaneous1}) together with $\Lambda_{\nu}$ are zero, or (\ref{simultaneous2}) together with $\Gamma_{\nu}$ are zero. Thus we have the following two cases: 

\noindent
{\bf Case I:} $\Lambda_{\nu} = 0$ ($\nu = 0,1$) and 
\begin{equation}
V_1(r)=-\frac{1}{2r^2}\left\{\gamma_1+\frac{3}{2}\gamma_2+3\gamma_2\log r\right\}+\gamma_3\,, \label{case1v1}
\end{equation}
where $\gamma_1$, $\gamma_2$ and $\gamma_3$ are constants.

\noindent
{\bf Case II:} $\Gamma_{\nu} = 0$ ($\nu = 0,1$) and 
\begin{equation}
V_1(r)=\frac{1}{2}r^2\chi_1-\frac{1}{2r^2}\chi_2+\chi_3\,, \label{case2v1}
\end{equation}
where $\chi_1$, $\chi_2$ and $\chi_3$ are constants.

Let us investigate these cases in detail.

\noindent
{\bf Case I:} From equations (\ref{sec2det1}) and (\ref{sec2det2}) together with (\ref{sec2det3}) we obtain the following forms of $\mathcal{F}_{1\nu}$ and $\mathcal{F}_{2\nu}$ ($\nu = 0,1$), which we present in polar form
\begin{eqnarray}
&\mathcal{F}_{1\nu} = \delta_{\nu, 1-\xi} \Big[\frac{1}{4}\Big((\mathcal{G}_{1\xi} \cos 2\theta + \mathcal{G}_{2\xi} 
\sin 2\theta)(2\gamma_1+3\gamma_2+4\gamma_3 r^2+6\gamma_2\log r) \nonumber \\
&+ 6 \mathcal{G}_{1\xi} \gamma_2 \log \frac{1}{r} \Big) - \mathcal{G}_{1\xi} r^2 \gamma_3 + 3 \mathcal{G}_{2\xi} \gamma_2 \theta\Big]
+ \gamma_{4\nu} r \sin \theta + \gamma_{5\nu} \,, \label{fullf1}  \\
&\mathcal{F}_{2\nu} = \delta_{\nu, 1-\xi} \Big[\frac{1}{4}\Big((\mathcal{G}_{1\xi} \sin 2\theta - \mathcal{G}_{2\xi} 
\cos 2\theta)(2\gamma_1+3\gamma_2+4\gamma_3 r^2+6\gamma_2\log r) \nonumber \\
&+ 6 \mathcal{G}_{2\xi} \gamma_2 \log \frac{1}{r} \Big) - \mathcal{G}_{2\xi} r^2 \gamma_3 - 3 \mathcal{G}_{1\xi} \gamma_2 \theta\Big]
- \gamma_{4\nu} r \cos \theta + \gamma_{6\nu}\,, 
 \label{fullg1} 
\end{eqnarray}
where $\gamma_{4\nu}$, $\gamma_{5\nu}$, $\gamma_{6\nu}$ are constants and $(\nu, \xi) = (0,1)$. Introducing (\ref{fullf1}) and (\ref{fullg1}) into the determining equations, obtained by equating the coefficients of the lower-order terms to zero in the commutativity equation, we obtain compatibility conditions for $\mathcal{F}_{3\nu}$ ($\nu = 0,1$) 
\begin{eqnarray}
\Bigg(16r^3(2V_{0,r}+rV_{0,rr})
+\Big(60 \gamma_1 \gamma_2 -4{\gamma_1}^2 + 63 {\gamma_2}^2-24r^2\gamma_2 \gamma_3-48r^4{\gamma_3}^2 \nonumber \\
+12\gamma_2\big(3 \gamma_2 (\log (\frac{1}{r})-(\log r - 5)\log r) - 2 \gamma_1 \log r \big)\Big)\Bigg) \nonumber \\
 + \Bigg(\frac{24\gamma_2 \gamma_{5\nu} \cos \theta + 72 \mathcal{G}_{2\nu} {\gamma_2}^2\theta \cos \theta +24\gamma_2 \gamma_{6\nu} \sin \theta - 72 \mathcal{G}_{1\nu} {\gamma_2}^2\theta \sin \theta}{\mathcal{G}_{1\nu} \cos \theta + \mathcal{G}_{2\nu} \sin \theta}\Bigg)=0\,.\label{combatibilityMo}
\end{eqnarray}

The invariance of potential implies that the term inside the last parenthesis in equation (\ref{combatibilityMo}) should be a constant, and from its form we see that it can only be the constant 0, which implies $\gamma_2=0$. After setting $\gamma_2=0$, $V_0$ can be obtained from (\ref{combatibilityMo}). However, setting $\gamma_2=0$ in (\ref{case1v1}) and considering (\ref{sectranspotsy}) we see that we can cancel the constant $\gamma_1$ by a gauge transformation. Hence, again we have a constant spin-orbit interaction potential. 

\noindent
{\bf Case II:} For this case equations (\ref{sec2det1}) and (\ref{sec2det2}) together with (\ref{sec2det3}) imply the following forms of $\mathcal{F}_{1\nu}$ and $\mathcal{F}_{2\nu}$ ($\nu = 0,1$),
\begin{eqnarray}
\mathcal{F}_{1\nu} = \delta_{\nu, 1-\xi} \Big[\frac{r^4\chi_1+\chi_2+2r^2\chi_3}{r}(\mathcal{G}_{3\xi} \sin \theta - \mathcal{G}_{4\xi} \cos \theta) - \frac{2 r^3 \chi_1}{3}(\mathcal{G}_{3\xi} \sin 3\theta - \mathcal{G}_{4\xi} \cos 3\theta) \Big] + \chi_{4\nu} r \sin \theta + \chi_{5\nu}\,, 
\nonumber \\
\mathcal{F}_{2\nu} = \delta_{\nu, 1-\xi} \Big[\frac{r^4\chi_1+\chi_2+2r^2\chi_3}{r}(\mathcal{G}_{3\xi} \cos \theta + \mathcal{G}_{4\xi} \sin \theta) + \frac{2 r^3 \chi_1}{3}(\mathcal{G}_{3\xi} \cos 3\theta + \mathcal{G}_{4\xi} \sin 3\theta) \Big] - \chi_{4\nu} r \cos \theta + \chi_{6\nu}\,,
\label{full2g1}
\end{eqnarray}
where $\chi_{4\nu}$, $\chi_{5\nu}$, $\chi_{6\nu}$ are constants and $(\nu, \xi) = (0,1)$. Introducing (\ref{full2g1}) into the determining equations, obtained by equating the coefficients of the lower-order terms to zero in the commutativity equation, we obtain compatibility conditions for $\mathcal{F}_{3\nu}$ ($\nu = 0,1$) 
\begin{eqnarray}
\Lambda_{\nu}\big(17r^8{\chi_1}^2+3{\chi_2}^2+36r^6\chi_1\chi_3+3r^3(V_{0,r} - r V_{0,rr})\big)
- 6r^5\chi_1 (\chi_{5\nu} \cos \theta - \chi_{6\nu} \sin \theta) = 0\,.
\label{case2compa}
\end{eqnarray}
From this compatibility condition (\ref{case2compa}) we conclude that we must have either $\chi_1=0$ or $\chi_{5\nu} = 0$ and $\chi_{6\nu} = 0$.

If $\chi_1=0$, then considering (\ref{sectranspotsy}) we can annihilate the constant $\chi_2$ in (\ref{case2v1}) by a gauge transformation and hence again we have a constant spin-orbit interaction potential.

If $\chi_{5\nu} = 0$ and $\chi_{6\nu} = 0$, then equation (\ref{case2compa}) implies 
\begin{equation}
V_0(r)=\frac{17\,r^6{\chi_1}^2}{72}+\frac{{\chi_2}^2}{8r^2}+\frac{3}{2}\,r^4\,\chi_1\chi_2+\frac{1}{2}r^2\epsilon_1\,,
\end{equation}
where $\epsilon_1$ is a constant. Upon introduction of this $V_0$ back into the determining equations coming from first- and zeroth-order terms, forces us to set $\chi_1=0$ in which case we are back in the previous case with a constant spin-orbit interaction potential.

\subsection{The potentials $V_0=\frac{y^2}{2}{V_1}^{\,2}(x)+F(x)$ and $V_1=V_1(x)$}
For these potentials the Hamiltonian (\ref{sec2eq1}) becomes, 
\begin{equation}
H=\frac{1}{2}({p_1}^2+{p_2}^2)+\frac{y^2}{2}{V_1}^{\,2}(x)+F(x)+V_1(x)\,\sigma_3\, L_3+\frac{1}{2}\sigma_3(L_3V_1(x))\,,
\label{sec3Hxdep}
\end{equation}
and the general second-order integral of motion to consider would be the one given in (\ref{generalintegralv3}). However, 
by making a similar analysis given in section 3.1, we see that the determining equations, obtained by equating the coefficients of the third-order terms to zero in the commutativity equation $[H, \tilde{X}_2] = 0$ forces us to write the general form of the second-order integral of motion as 
\begin{eqnarray}
X_2 = \mathcal{G}_1 (L_3\,p_1+p_1\,L_3) + \mathcal{G}_2 (L_3\,p_2+p_2\,L_3) + \mathcal{G}_3 ({p_1}^2-{p_2}^2) + 2\, \mathcal{G}_4\,p_1\,p_2 + \mathcal{G}_5\,L_3^2 + X_1\,,
\label{secondorderintmotionxdep}
\end{eqnarray}
where $\mathcal{G}_{\tau}$ ($\tau = 1, \dots, 5$) are constants and 
$X_1$ is the first-order integral of motion given in (\ref{integralofmotion}). These are considered as
\begin{equation}
\mathcal{G}_{\tau} = \mathcal{G}_{\tau 0} + \mathcal{G}_{\tau 1}\, \sigma_3\,, \quad  \tau = 1,\ldots,5\,, \qquad \mathcal{F}_{\mu} = \mathcal{F}_{\mu 0} + \mathcal{F}_{\mu 1}\, \sigma_3\,, \quad  \mu = 1,2,3\,.
\end{equation}
Notice that a term proportional to $L_3^2$ is present in (\ref{secondorderintmotionxdep}), since it does not commute with the Hamiltonian given in (\ref{sec3Hxdep}).

The determining equations, obtained by equating the coefficients
of the second-order terms to zero in the commutativity equation $[H, X_2] = 0$, read  
\begin{eqnarray}
&2 \sigma_3 \big((\mathcal{G}_4 - \mathcal{G}_2 y) V_1 + y (\mathcal{G}_3 - 2 \mathcal{G}_1 y + \mathcal{G}_5 y^2) V_{1,x}\big) + \mathcal{F}_{1,x} = 0\,,  
\label{type2pot1} \\
&2 \sigma_3 \big((\mathcal{G}_4 + \mathcal{G}_1 x)V_1 + x (\mathcal{G}_4 + \mathcal{G}_1 x - (\mathcal{G}_2 + \mathcal{G}_5 x) y) V_{1,x}\big) - \mathcal{F}_{2,y} = 0\,,  
\label{type2pot2} \\
&2 \sigma_3 \big((2 \mathcal{G}_3 - \mathcal{G}_2 x - \mathcal{G}_1 y) V_1 + (\mathcal{G}_3 x - (\mathcal{G}_4 + 3 \mathcal{G}_1 x) y + (\mathcal{G}_2 + 2 \mathcal{G}_5 x) y^2) V_{1,x}\big) - \mathcal{F}_{1,y} - \mathcal{F}_{2,x} = 0\,.  
\label{type2pot3}
\end{eqnarray}
From equations (\ref{type2pot1})--(\ref{type2pot3}) we obtain compatibility conditions for $\mathcal{F}_{1\nu}$ and $\mathcal{F}_{2\nu}$ ($\nu = 0,1$), 
\begin{eqnarray}
12 (\mathcal{G}_1 - \mathcal{G}_5 y) V_{1,x} + 4 \big(\mathcal{G}_4 + 2 \mathcal{G}_1 x - (\mathcal{G}_2 + 2 \mathcal{G}_5 x) y\big) V_{1,xx} + x \big(\mathcal{G}_4 + \mathcal{G}_1 x - (\mathcal{G}_2 + \mathcal{G}_5 x) y\big) V_{1,xxx} = 0\,. 
\label{type2com}
\end{eqnarray}
Since $V_1$ is a function of $x$ only, the coefficients of $y$ in (\ref{type2com}) must vanish separately (\emph{i.e.}, either $\mathcal{G}_1 = 0$ and  $\mathcal{G}_4 = 0$ or $\mathcal{G}_2 = 0$ and  $\mathcal{G}_5 = 0$). If $\mathcal{G}_1 = 0$ and  $\mathcal{G}_4 = 0$, then (\ref{type2com}) implies 
\begin{eqnarray}
12 \mathcal{G}_{5\nu} V_{1,x} + 4 \big(\mathcal{G}_{2\nu} + 2 \mathcal{G}_{5\nu} x\big) V_{1,xx} + x \big(\mathcal{G}_{2\nu} + \mathcal{G}_{5\nu} x) V_{1,xxx} = 0\,, \quad \nu = 0,1\,.
\label{type2cominsepform1}
\end{eqnarray}
One of these equations, say the one with $\nu = 0$, can be solved for $V_1$ and upon introducing this solution into the other equation we obtain the following constraint on the constants 
\begin{eqnarray}
\frac{\mathcal{G}_{21}}{\mathcal{G}_{20}} = 
\frac{\mathcal{G}_{51}}{\mathcal{G}_{50}} = \lambda_1\,,
\label{constarintforconstants1}
\end{eqnarray}
where $\lambda_1$ is a constant. On the other hand if $\mathcal{G}_2 = 0$ and  $\mathcal{G}_5 = 0$, then (\ref{type2com}) implies
\begin{eqnarray}
12 \mathcal{G}_{1\nu} V_{1,x} + 4 \big(\mathcal{G}_{4\nu} + 2 \mathcal{G}_{1\nu} x\big) V_{1,xx} + x \big(\mathcal{G}_{4\nu} + \mathcal{G}_{1\nu} x) V_{1,xxx} = 0\,, \quad \nu = 0,1\,.
\label{type2cominsepform2}
\end{eqnarray}
Similarly, solving (\ref{type2cominsepform2}) for $V_1$ gives the following constraint on the constants 
\begin{eqnarray}
\frac{\mathcal{G}_{11}}{\mathcal{G}_{10}} = 
\frac{\mathcal{G}_{41}}{\mathcal{G}_{40}} = \lambda_2\,,
\label{constarintforconstants2}
\end{eqnarray}
where $\lambda_2$ is a constant.

In both cases, the compatibility conditions for $\mathcal{F}_{1\nu}$ and $\mathcal{F}_{2\nu}$ ($\nu = 0,1$), imply the following generic form of the interaction potential 
\begin{eqnarray}
V_1(x)=\frac{-3 x^2 \alpha_1 / \mathcal{G}_{\mu 0}^2 + (2 x + \lambda_a){\alpha}_2}{6 x^2 (x + \lambda_a)^2} + \alpha_3\,, \quad a = 1,2\,, 
\label{genericpotforxpot}
\end{eqnarray}
where $\alpha_{1}$, $\alpha_{2}$, $\alpha_{3}$, $\lambda_a$ are constants and $\mu$ is either $1$ or $4$, or $2$ or $5$. 

Only exception to this generic potential occurs if we have $\mathcal{G}_{\mu 0} = 0$, in which case the interaction potential $V_1$ becomes 
\begin{eqnarray}
V_1(x)=\frac{\zeta_1}{6 x^2} + x \zeta_2 + \zeta_3\,,
\label{exceptiontogenericpot}
\end{eqnarray}
where $\zeta_{1}$, $\zeta_{2}$ and $\zeta_{3}$ are constants.

In all three cases we proceed in a similar fashion as we did for rotationally invariant potentials. More specifically, for two types of potentials (\ref{genericpotforxpot}) and (\ref{exceptiontogenericpot}) we find $\mathcal{F}_{1 \nu}$ and $\mathcal{F}_{2 \nu}$ ($\nu = 0,1$) from equations (\ref{type2pot1}) and (\ref{type2pot2}) together with (\ref{type2pot3}). Then introducing these forms of $\mathcal{F}_{1 \nu}$ and $\mathcal{F}_{2 \nu}$ ($\nu = 0,1$) into the determining equations, obtained by equating the coefficients of the lower-order terms to zero in the commutativity equation, we obtain compatibility conditions for $\mathcal{F}_{3\nu}$ ($\nu = 0,1$). In order to satisfy these compatibility conditions, either $\zeta_2$ must vanish in (\ref{exceptiontogenericpot}) or for the generic potential (\ref{genericpotforxpot}) $\mathcal{G}_{5}$ must also vanish for the case $\mathcal{G}_{1} = \mathcal{G}_{4} = 0$ and $\mathcal{G}_{4}$ must vanish for the case $\mathcal{G}_{2} = \mathcal{G}_{5} = 0$. Unfortunately, these equations are rather long to present here.    

If $\zeta_2 = 0$ in (\ref{exceptiontogenericpot}), then by means of gauge transformation (\ref{sectranspotsy}) we can annihilate the constant  $\zeta_1 = 0$ in (\ref{exceptiontogenericpot}) and have a constant spin-orbit interaction potential. If $\mathcal{G}_{5} = 0$ in addition to $\mathcal{G}_{1} = \mathcal{G}_{4} = 0$ or $\mathcal{G}_{4} = 0$ in addition to $\mathcal{G}_{2} = \mathcal{G}_{5} = 0$, then we repeat the analysis from the beginning and find the following results 

\noindent
{\bf Case I:} $\mathcal{G}_{1} = 0$, $\mathcal{G}_{4} = 0$ and $\mathcal{G}_{5} = 0$
\begin{eqnarray}
V_1(x) = \frac{\alpha_1}{6x^2}+\alpha_3\,, \quad
F(x) = \frac{{\alpha_1}^2}{72x^2}+\frac{x^2{\alpha_3}^2}{2}\,, \quad V_0(x,y) = \frac{1}{72}\left(\frac{{\alpha_1}^2}{x^2}+36x^2{\alpha_3}^2+\frac{y^2(\alpha_1+6x^2\alpha_3)^2}{x^4}\right)\,, 
\label{sec32res1}
\end{eqnarray}
where $\alpha_{1}$, $\alpha_{2}$ and $\alpha_{3}$ are constants.

\noindent
{\bf Case II:} $\mathcal{G}_{2} = 0$, $\mathcal{G}_{5} = 0$ and $\mathcal{G}_{4} = 0$
\begin{eqnarray}
V_1(x) = -\frac{\alpha_2}{2x^2}+\alpha_3\,, \quad 
F(x) = \frac{{\alpha_2}^2}{8x^2}+\frac{x^2{\alpha_3}^2}{2}\,, \quad V_0(x,y) = \frac{1}{8}\left(\frac{{\alpha_2}^2}{x^2}+4x^2{\alpha_3}^2+4 y^2\big(\alpha_3 -\frac{\alpha_2}{2 x^2}\big)^2 \right)\,, 
\label{sec32res2}
\end{eqnarray}
where again $\alpha_{1}$, $\alpha_{2}$ and $\alpha_{3}$ are constants.

Notice that the potentials $V_1(x)$ and $V_0(x,y)$ given in (\ref{sec32res1}) and  (\ref{sec32res2}) are exactly the same ($\alpha_2\rightarrow -\frac{1}{3}\alpha_1$). 

We conclude that for these two cases once again we have a constant spin-orbit interaction potential (up to the gauge transformation (\ref{sectranspotsy})) and hence all the second-order integrals of motion that are obtained from the analysis can be expressed in terms of the first-order ones given in (\ref{sec2algebra1}).

\section{Conclusions}
The main result of this paper can be given as a theorem. 

\noindent
{\bf Theorem 1.} In Euclidean plane $E_2$, for Hamiltonians of the type (\ref{sec2eq1}) admitting a first-order integral, any second-order integral can necessarily be expressed as a combination of first-order integrals. Or in particular, there exists no nontrivial second-order integrals of motion of the form (\ref{generalintegralv3}) for the integrable Hamiltonian systems (\ref{sec3H}) and (\ref{sec3Hxdep}).

This result, which is valid for the generic Hamiltonian systems of the type (\ref{sec2eq1}), proves that no nontrivial generic second-order integrals of motion exist and hence carries one step further the systematic study of the classification of integrable and superintegrable Hamiltonian systems involving spin in Euclidean plane $E_2$.

In an earlier article \cite{WinYur1} it was shown that in the presence of spin first-order integrable and superintegrable systems exist in $E_2$. The superintegrable Hamiltonian of such systems allows the separation of variables in polar and Cartesian coordinates. Indeed, the Pauli-Schr\"odinger equation for them can be exactly solved. The integrable Hamiltonians also allow the separation of variables in polar and Cartesian coordinates. However, in order to solve them exactly the interaction potential $V_1(r)$ and scalar potential $V_0(r)$ have to be specified. For instance, choosing $V_1(r) = \lambda/r^2$ ($\lambda =$ constant) and $V_0(r) = \alpha r^2/2$ ($\alpha =$ constant), the radial part of the wave function of the Pauli-Schr\"odinger equation can be expressed in terms of Laguerre polynomials. 
 
Another way of dealing such problems is to search for potentials admitting an additional integral of motion. Since it was shown in \cite{WinYur1} that there exist exactly one first-order integral of motion for these integrable systems (see the equations (\ref{rotinvsys}) and (\ref{xdeppotsys})), the additional integral of motion should be higher-order one. In this article we search for the second-order ones and sum up our results as Theorem 1. In a future work third- and higher-order integrals of motion can be investigated for such systems. 

Investigation of integrability and superintegrability properties of other type of Hamiltonian systems involving spin in Euclidean plane is in progress.

\section{Acknowledgments}
The author thanks Pavel Winternitz and the anonymous reviewers for their valuable comments and suggestions to improve the quality of the paper. This work is partially supported by the Scientific and Technological Research Council of Turkey (T\"{U}B\.{I}TAK).



\end{document}